\documentclass[aps,prl,preprint,superscriptaddress]{revtex4-2}

\usepackage{graphicx}
\usepackage{natbib,hyperref}
\usepackage[english]{babel}

\begin{document}

\title{Coupling of the triple-Q state to the atomic lattice by anisotropic symmetric exchange}

\author{Felix Nickel}
\email[Email: ]{nickel@physik.uni-kiel.de}
\affiliation{Institut f\"ur Theoretische Physik und Astrophysik, Christian-Albrechts-Universit\"at zu Kiel, D-24098 Kiel, Germany}
\author{Andr\'e Kubetzka}
\affiliation{Department of Physics, University of Hamburg, Jungiusstraße 11, 20355 Hamburg, Germany}
\author{Soumyajyoti Haldar}
\affiliation{Institut f\"ur Theoretische Physik und Astrophysik, Christian-Albrechts-Universit\"at zu Kiel, D-24098 Kiel, Germany}
\author{Roland~Wiesendanger}
\affiliation{Department of Physics, University of Hamburg, Jungiusstraße 11, 20355 Hamburg, Germany}
\author{Stefan Heinze}
\affiliation{Institut f\"ur Theoretische Physik und Astrophysik, Christian-Albrechts-Universit\"at zu Kiel, D-24098 Kiel, Germany}
\affiliation{Kiel Nano, Surface, and Interface Science (KiNSIS), University of Kiel, Germany}
\author{Kirsten von Bergmann}
\email{kirsten.von.bergmann@physik.uni-hamburg.de}
\affiliation{Department of Physics, University of Hamburg, Jungiusstraße 11, 20355 Hamburg, Germany}

\date{\today}

\begin{abstract}
We identify the triple-Q (3Q) state as magnetic ground state in Pd/Mn and Rh/Mn bilayers on Re(0001) using spin-polarized scanning tunneling microscopy and density functional theory. An atomistic model reveals that in general the 3Q state with tetrahedral magnetic order and zero net spin moment is coupled to a hexagonal atomic lattice in a highly symmetric orientation via the anisotropic symmetric exchange interaction, whereas other spin-orbit coupling terms cancel due to symmetry. Our experiments are in agreement with the predicted orientation of the 3Q state. A distortion from the ideal tetrahedral angles leads to other orientations of the 3Q state which, however, results in a reduced topological orbital magnetization compared to the ideal 3Q state.
\end{abstract}

\maketitle

\begin{figure}[b]
\includegraphics[width=0.5\linewidth]{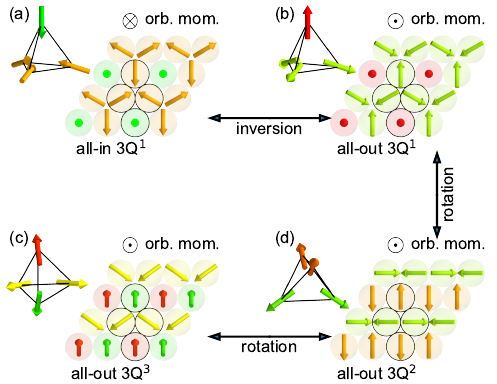}
\caption{Illustration of the 3Q state in different highly symmetric orientations with respect to the hexagonal atomic lattice; the respective direction of the topological orbital moment is indicated. Note that the 3Q states in (a) and (b) are connected by an inversion of all spins, whereas the states in (b-d) can be converted into each other by a continuous rotation of the spin structure.}
\label{fig:intro}
\end{figure}

The triple-Q (3Q) state is a three-dimensional spin structure on a two-dimensional hexagonal lattice. It can be understood as a superposition of three symmetry equivalent spin spiral (1Q) states resulting in a non-coplanar magnetic state with tetrahedral angles between all adjacent magnetic moments and four atoms in the magnetic unit cell~(Fig.~\ref{fig:intro}). This fascinating magnetic state was predicted more than 20 years ago~\cite{Momoi1997,Kurz2001}, but the first experimental observation was reported only recently for an hcp-stacked Mn monolayer on Re(0001) using spin-polarized scanning tunneling microscopy (SP-STM)~\cite{Spethmann2020}. The ideal 3Q state does not exhibit a net spin moment, however, theoretical investigations have shown a significant topological orbital magnetization and a spontaneous topological Hall effect, without the necessity of spin-orbit coupling~\cite{Martin2008,Hoffmann2015,Hanke2016,Grytsiuk2020}. Recently, neutron scattering experiments on the layered material Co$_{1/3}$TaS$_2$ were interpreted as validation of the 3Q state, and indeed, transport measurements of this bulk system are in agreement with a magnetic-field-induced switching of the orbital moment direction and reveal the topological Hall effect~\cite{Park2023,Takagi2023}. For magnetic 3Q states in a two-dimensional system the orbital moment is always perpendicular to the layer and the two different directions are related to the sign of the scalar spin chirality in an \emph{all-in} versus \emph{all-out} configuration (see Fig.~\ref{fig:intro}). Another emergent phenomenon of the 3Q state is highlighted by a theoretical proposal that demonstrates a topological superconducting phase induced by this non-coplanar magnetic state when it is adjacent to a conventional superconductor~\cite{Bedow2020}.

The 3Q state can arise in frustrated antiferromagnets, for instance as a ground state in a hexagonal lattice of spins with antiferromagnetic nearest and next-nearest neighbor interactions ($1 < J_1 / J_2 < 8$)~\cite{Kurz2001,Spethmann2020}. To lower the energy with respect to the otherwise degenerate 1Q state higher-order interactions (HOI) are necessary. One prominent HOI term is the biquadratic interaction, which is one of the four-spin interactions which arise in fourth order in a perturbative expansion of the Hubbard model~\cite{Takahashi1977,MacDonald1988,Hoffmann2020}. The aforementioned topological orbital moments that can occur in these non-coplanar magnetic states can interact with the emergent magnetic field leading to the topological chiral magnetic interaction that constitutes a sixth order term~\cite{Grytsiuk2020}. Due to this topological chiral magnetic interaction large distortions from the perfect tetrahedron angle can occur as proposed for Mn/Re(0001)~\cite{Haldar2021}.

The possibility of a coupling of the ideal 3Q state to the crystal lattice is an intriguing question which has not been addressed so far. Figure~\ref{fig:intro} displays several 3Q states which have different orientations of the spins with respect to the magnetic lattice plane: a 3Q state with a given orbital moment can occur in three symmetric spin orientations with respect to the plane, denoted as 3Q$^1$, 3Q$^2$, and 3Q$^3$~\cite{Spethmann2020}. An energy variation between these states, i.e.\ a preferred coupling to the lattice, must originate from spin-orbit interaction. The most prominent spin-orbit interaction term is the magneto-crystalline anisotropy energy (MAE). However, for the ideal 3Q state with its 4 spins arranged in a tetrahedral fashion the leading MAE term vanishes. In non-collinear magnetic states often the spin-orbit coupling induced Dzyaloshinskii-Moriya interaction (DMI) contributes, but in the 3Q state for symmetry reasons also this term cancels (because the two spins on opposite sides of any reference spin have the same magnetization direction). A further interaction that originates from spin-orbit coupling is the anisotropic symmetric exchange interaction (ASE), also referred to as pseudo-dipolar interaction or compass anisotropy~\cite{Smith1976,Staunton1988}. It has been shown to play a role for the coupling of the spins to the lattice in a uniaxial antiferromagnetic state~\cite{Spethmann2020}, and it can induce skyrmion lattices in two-dimensional van der Waals magnets~\cite{Amoroso2020} and in centroysymmetric crystals lacking DMI~\cite{Hirschberger2021}. Recently another spin-orbit coupling induced term has been proposed, namely the spin-chiral interaction (SCI) which results from the coupling of a spin to the topological orbital moment arising in non-coplanar states~\cite{Brinker2019,Szunyogh2019,Mankovky2020,Grytsiuk2020}.

Here, we use SP-STM to investigate the magnetic ground states of Pd/Mn and Rh/Mn bilayers on Re(0001) in real space. We find that both systems exhibit a hexagonal magnetic superstructure, indicative of the 3Q state with tetrahedral arrangement of the magnetic moments. Based on density functional theory (DFT) we confirm the 3Q magnetic ground state of these two systems and calculate the resulting topological orbital moments. We show that the ASE couples the state to the atomic lattice resulting for both systems in the 3Q$^1$ configuration, in which one of the four spins of the tetrahedron points perpendicular to the surface, Fig.~\ref{fig:intro}(a,b). In addition, the ASE also determines the direction of the in-plane magnetization components within this 3Q$^1$ configuration, selecting a group of very specific states out of the large number of possible 3Q configurations.

\begin{figure}
\includegraphics[width=0.5\linewidth]{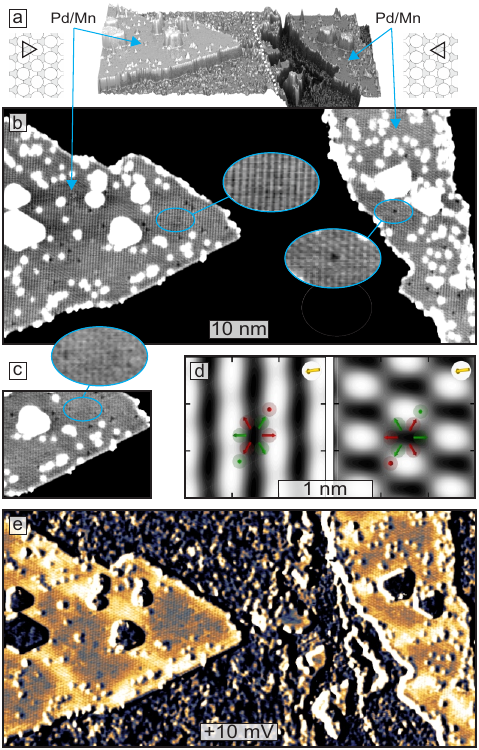}
\caption{(a)~Perspective view of a sample of about 0.3 atomic layers of Pd on about 0.8 atomic layers of Mn on Re(0001); the dotted white line marks a buried Re(0001) step edge. (b)~Spin-resolved constant-current STM image of the same area as shown in (a), with the height contrast adjusted separately for the two Pd/Mn bilayer areas to $\delta z = \pm 10$\,pm ($U=+10$\,mV, $I=5$\,nA, $T=4.2$\,K, Cr-bulk tip). (c)~Selected area of (b) imaged with a different tip magnetization direction, exhibiting a different pattern. (d)~SP-STM images of the 3Q$^1$ ground state of Pd\textsubscript{fcc}/Mn\textsubscript{fcc}/Re(0001) calculated from DFT based on the spin-polarized model of STM \cite{Wortmann2001}. The images were calculated for a bias voltage of $-100 \, \text{meV}$ and at a distance of $6$ {\AA} above the surface. The tip magnetization direction is indicated by the orange arrow. The green and red spheres with arrows denote the Pd and Mn atoms, respectively, and their magnetic moment directions. The two images represent the same 3Q rotated by $180^{\circ}$, as realized in two Pd/Mn bilayers on adjacent terraces. (e)~Map of differential tunnel conductance acquired simultaneously to (b).}
\label{fig:exp_spstm_PdMnRe}
\end{figure}

The starting point of our SP-STM investigation is the Mn monolayer on Re(0001), which preferentially grows in fcc stacking. The experimentally found magnetic ground state is the row-wise antiferromagnetic state that is the constituing 1Q state of the tetrahedral 3Q state~\cite{Spethmann2020}, however, DFT predicts that the 1Q and the 3Q states are nearly degenerate. To drive the system into a 3Q ground state we try to tune the interactions by covering the extended fcc-Mn monolayer by non-magnetic overlayers~\cite{Romming2013,romming18}, namely Pd and Rh. When submonolayer amounts of Pd or Rh are deposited we observe pseudomorphic growth of the overlayer on top of the fcc-stacked Mn monolayer and a decoration at the Mn step edges, see perspective view of a Pd/Mn/Re(0001) sample in Fig.~\ref{fig:exp_spstm_PdMnRe}(a); note that due to the hcp crystal structure of Re(0001) the relative layer positions of the Pd and Mn layers are interchanged on adjacent Re terraces, cf.\ the two sketches of the bilayer structures, which are rotated by $180^{\circ}$ with respect to each other. Analysis of the step edge directions of the overlayer islands reveals that Pd grows in only one stacking on the fcc-Mn monolayer, with small patches of Pd islands on top of the Pd/Mn bilayer. In contrast, the Rh monolayer occurs in both possible stackings on top of the fcc-Mn/Re(0001) (see Supplemental Material~\cite{supplmat}). 

In Fig.~\ref{fig:exp_spstm_PdMnRe}(b) the height contrast for the two Pd/Mn bilayer areas is adjusted separately to the same value for better visibility. We find that the Pd/Mn exhibits an atomic-scale pattern of magnetic origin, which slightly varies over the islands but dominantly appears as vertical stripes on the left island, and like a hexagonal pattern on the right island, see enlarged insets. The length scale of the magnetic superstructure indicates that the 3Q state is the magnetic ground state. Depending on the tip magnetization direction relative to the 3Q state, different superstructure patterns are expected; indeed, in a measurement with a different tip magnetization direction the left Pd/Mn island can also exhibit a hexagonal pattern, see Fig.~\ref{fig:exp_spstm_PdMnRe}(c). Also the Rh/Mn bilayers show the same magnetic superstructure unit cell for both Rh stackings~\cite{supplmat} albeit with very small contrast amplitude. Nevertheless, this is an indication that also the Rh/Mn bilayer exhibits the 3Q state. With the help of SP-STM simulations we can find a combination of 3Q state and tip magnetization that reproduces the magnetic patterns observed on the two Pd/Mn areas in the measurement of Fig.~\ref{fig:exp_spstm_PdMnRe}(b). We use the constraint that the configuration of a bilayer, and thus also a given 3Q state, must be rotated by $180^{\circ}$ on adjacent Re terraces and find a canted tip magnetization that nicely reproduces the experimentally observed magnetic contrast, see the DFT-based SP-STM simulations displayed in (d). Note that an inversion of all sample spins (or the tip magnetization direction), only slightly changes the observed patterns.

The d$I$/d$U$~map in Fig.~\ref{fig:exp_spstm_PdMnRe}(e) is acquired simultaneously to (b). It shows a signal variation that is spatially correlated to the slight changes of the magnetic pattern on the Pd/Mn islands, suggesting a contribution of an electronic contrast mechanism due to local changes in the magnetic state~\cite{Bergmann2012,Hanneken:15.1}. We interpret the darker areas as magnetic domains, and the brighter areas as transition regions between them. Indeed, close analysis shows that there is a phase shift between different areas of the stripe contrast in the left island~\cite{supplmat}. In contrast to the sharp domain walls previously observed for the 3Q state in hcp-stacked Mn on Re(0001)~\cite{Spethmann2020}, the transitions between different 3Q areas in Pd/Mn are much more extended. It is worth noting that, despite the presence of several independent domains, no rotational domains are observed. For both the 3Q$^2$ and 3Q$^3$ state rotational domains are expected to coexist, and with an arbitrary and unknown tip magnetization direction, as in our case, they could be easily discriminated. This suggests that we have the 3Q$^1$ configuration in our Pd/Mn bilayer. 

\begin{figure}
\includegraphics[width=0.5\linewidth]{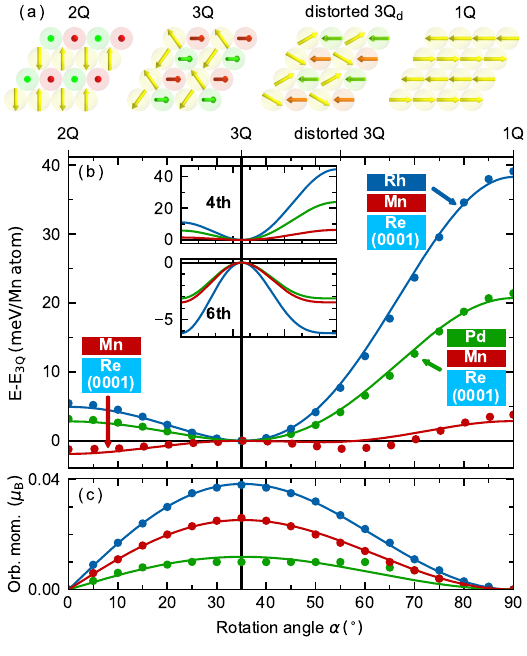}
\caption{(a) Sketch of the 2Q, the 3Q, a distorted 3Q and the 1Q (RW-AFM) state. (b) Energy dispersion for spin states along the path 2Q-3Q-1Q according to Eq.~(\ref{eq:2Q3Q1Q}) for Pd/Mn/Re(0001), Rh/Mn/Re(0001) and hcp-stacked Mn/Re(0001). The symbols represent DFT data, the lines show fits to the DFT values using HOI contributions of 4th and 6th order. The insets show the contributions of the 4th and 6th order separately. (c) Absolute value of the topological orbital moment of the Mn atoms in the films. Symbols show DFT values and lines a fit to the scalar spin chirality. Data for Mn/Re(0001) are taken from Ref.~\cite{Haldar2021}.}
\label{fig:XMnRe_231Q}
\end{figure}

To understand the experimental observations we have performed DFT calculations for Pd/Mn and Rh/Mn bilayers on Re(0001) based on the full-potential linearized augmented plane-wave method as implemented in the {\tt FLEUR} code \cite{FLEUR,Kurz2004,Heide2009} and using the projector augmented wave method as implemented in the {\tt VASP} code \cite{VASP1,VASP2} (see Supplemental Material for computational details). In order to scan a large part of the phase space of magnetic states we have first calculated the total energy of spin spiral (1Q) states which are the fundamental solution of the classical Heisenberg model and allow to calculate the isotropic pair-wise exchange constants \footnote{Note, that the DMI does not play a role for the magnetic ground state in these films. The contribution of spin-orbit coupling (SOC) to the energy dispersion of spin spirals, which allows the determination of the DMI parameters, is displayed in the Supplemental Material. \cite{supplmat}}. These calculations show that out of all 1Q states the row-wise antiferromagnetic (RW-AFM) state (Fig.~\ref{fig:XMnRe_231Q}(a)) and the N\'eel state are the lowest 1Q state for Pd/Mn/Re(0001) and Rh/Mn/Re(0001), respectively~\cite{supplmat}. However, if also superposition states that can be stabilized by HOI are considered our DFT calculations show that the 3Q state (Fig.~\ref{fig:intro}) is by about 20 meV/Mn atom lower than the lowest 1Q state for both systems. This is in agreement with the experimentally observed $p(2 \times 2)$ magnetic unit cell and we conclude that the 3Q state is the ground state for both the Pd/Mn and the Rh/Mn bilayer on Re(0001).

The 1Q state and its corresponding 3Q superposition state are degenerate in energy within the Heisenberg model \cite{Haldar2021}, and the large total energy difference obtained by DFT must stem from HOIs. In order to evaluate the role of the HOIs for these systems we calculate the energy of different superposition states along the path continuously connecting the 1Q and the 2Q state via the 3Q state, see Fig.~\ref{fig:XMnRe_231Q}~\cite{Haldar2021}.

The configuration of the superposition state is given by
\begin{equation}
    \mathbf{s}_i (\alpha)= \mathbf{s}_i^{\rm 2Q} \cos(\alpha) +  \mathbf{s}_i^{\rm 1Q} \sin(\alpha),
    \label{eq:2Q3Q1Q}
\end{equation}
where $\mathbf{s}_i (\alpha)$ represents a spin at lattice site $i$ in the state characterized by $\alpha$, and $\mathbf{s}_i^{\rm 2Q}$ and $\mathbf{s}_i^{\rm 1Q}$ are the spin orientations for the 2Q and 1Q state, respectively (Fig.~\ref{fig:XMnRe_231Q}(a), see also Ref.~\cite{supplmat}). The angle $\alpha$ is varied between $0^{\circ}$ (2Q state) and $90^{\circ}$ (1Q state). The 3Q state occurs at an angle of $\arcsin(1/\sqrt{3}) \approx 35.26 ^{\circ}$. 

The energy difference between the 3Q and the 1Q state (RW-AFM), $\Delta E_{\rm 3Q-1Q}$, is on the order of 20 and 40 meV/Mn atom for the Pd/Mn and the Rh/Mn bilayer, respectively. The DFT total energy differences along the path can be fitted by HOI contributions of 4th and 6th order (Fig.~\ref{fig:XMnRe_231Q}(b) and insets). We find that while $\Delta E_{\rm 3Q-1Q}$ for both systems is dominated by the 4th order interaction, also the 6th order interaction is significant with a value of about $15\%$ compared to the 4th order contribution. As a reference, we also show the energies for the hcp-Mn/Re(0001), which was found to exhibit the 3Q state (note that in the bilayers studied here the Mn is in fcc-stacking). We find that the energy variation along the 2Q-3Q-1Q path is significantly smaller; here the contributions of 4th and 6th order terms are of similar size, giving rise to a minimum at an angle of about $55^{\circ}$, which can be interpreted as a distorted 3Q state~\cite{Haldar2021}. One can also calculate the HOI constants explicitly \cite{Hoffmann2020}, see Supplemental Material 
for the values.

\begin{figure}
\includegraphics[width=0.5\linewidth]{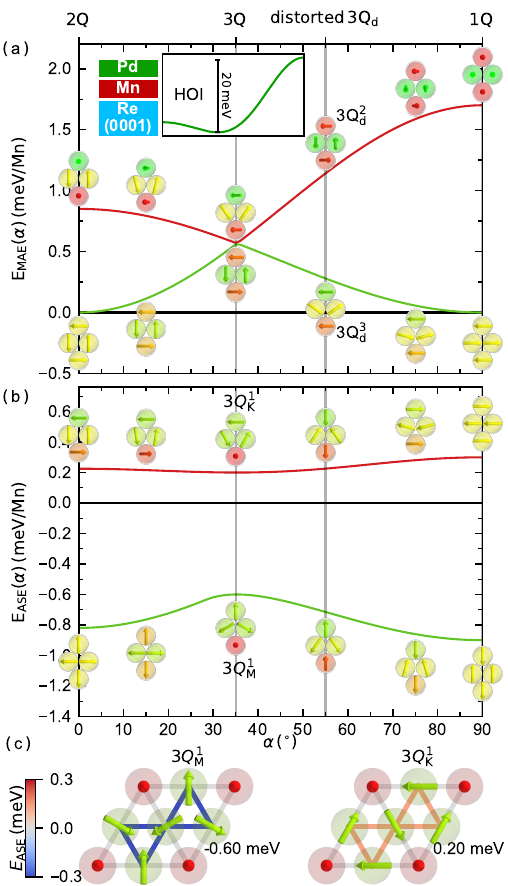}
\caption{Energy contributions of (a) the magnetocrystalline anisotropy energy (MAE) and (b)  the anisotropic symmetric exchange (ASE) for spin states along the path 2Q-3Q-1Q according to Eq.~(\ref{eq:2Q3Q1Q}) with interaction strengths obtained via DFT for Pd/Mn/Re(0001). The line indicates the minimal (green) and maximal (red) energy contributions out of all possible global spin rotations. For selected points along the path the spin configurations that minimize or maximize the energy are displayed. (c)~Sketches of the two 3Q states with minimal and maximal ASE energy, the bonds are colorized according to their individual contributions to the total ASE. In the 3Q$^{1}_{\textrm{M}}$ state all in-plane components are along the $\overline{\Gamma \textrm{K}}$ direction, whereas in the 3Q$^{1}_{\textrm{K}}$ state they are along $\overline{\Gamma \textrm{K}}$, i.e.\ the close packed atomic rows.} 
\label{fig:theo_MAE_ASE}
\end{figure}

The 6th order terms have the largest effective value for Rh/Mn/Re(0001). This is accompanied by a large topological orbital moment (Fig.~\ref{fig:XMnRe_231Q}(c)), which maximizes for the 3Q state, and decreases if the 3Q state is distorted towards the 2Q or the 1Q state. The direction of the orbital moment is linked to the specific spin configuration and inverts between the all-in and the all-out states; however, the continuous rotation of all spin directions does not alter the size or the sign of the orbital moment (cf. Fig.~\ref{fig:intro}). The presence of topological orbital moments, which occur due to the scalar spin chirality $\chi_{ijk}=\mathbf{s}_i \cdot (\mathbf{s}_j \times \mathbf{s}_k)$ even in the absence of spin-orbit coupling (SOC), is an indicator for topological-chiral interactions \cite{Grytsiuk2020,Haldar2021}. 

We have shown that the ideal undistorted 3Q state is the ground state of Pd/Mn and Rh/Mn bilayers on Re(0001). Now we want to analyze the impact of SOC onto the specific configuration of the 3Q state. In Fig.~\ref{fig:theo_MAE_ASE} we show the energy due to MAE and ASE along the 2Q-3Q-1Q path for the Pd/Mn bilayer \footnote{Note that the spin-chiral interaction which also arises due to SOC \cite{Grytsiuk2020} does not play a role for the coupling of the 3Q state to the atomic lattice as can be shown analytically \cite{supplmat}.}. For each state along the path, we allowed global rotations of the complete spin structure. Then we performed a global optimization to find the spin orientations of lowest (green) and highest (red) energy. From our DFT calculations we found that the MAE prefers an in-plane orientation of the magnetic moments with a strength of $-1.70 \, \text{meV/Mn atom}$ compared to an out-of-plane alignment. The 2Q state and the 1Q state have configurations with all spins in the easy plane and accordingly there is a selection of the specific configuration by the MAE, with an energy difference of 0.7 and 1.7 meV/Mn atom between different configurations for the 2Q and 1Q, respectively. In contrast, an undistorted 3Q state with tetrahedral angles between all nearest neighbor moments cannot gain MAE (Fig.~\ref{fig:theo_MAE_ASE}(a)), as obvious from symmetry considerations. Therefore, the 3Q always has the higher MAE compared to the 2Q or the 1Q state, for both easy plane and easy axis systems (by $0.7$ and $0.2-1$ meV/Mn atom, respectively). However, for Pd/Mn the large HOI contribution favoring the 3Q state (see inset) more than overcompensates the MAE between the 3Q and the 2Q/1Q states. In contrast, when a distorted 3Q state is favored by the 6th order HOIs, as is the case for hcp-Mn/Re(0001) with $\alpha \approx 55^{\circ}$, the MAE can select between different orientations of spin configurations.

Figure~\ref{fig:theo_MAE_ASE}(b) shows the same graph for the ASE energy, which can be calculated for a given magnetic state via
\begin{equation}
    E_{\rm ASE} = - \sum_{i, j} J_{\rm ASE} (\mathbf{s}_i \cdot \mathbf{d}_{ij})(\mathbf{s}_j \cdot \mathbf{d}_{ij}) \quad .
\end{equation}
Here $\mathbf{d}_{ij}$ is the normalized connection vector between two lattice sites defined by $i, j$ and $\mathbf{s}_i, \mathbf{s}_j$ are the corresponding normalized spin moments. This interaction favours either a ferromagnetic or an antiferromagnetic alignment along the connection $\mathbf{d}_{ij}$ depending on the sign of $J_{\rm ASE}$. If one or both magnetic moments are perpendicular to $\mathbf{d}_{ij}$ the contribution vanishes. For Pd/Mn the value of the ASE calculated by DFT is $J_{\rm ASE}=-0.30 \, \text{meV/Mn atom}$.

In contrast to the MAE, the ASE favors a specific orientation of the spin structure with respect to the lattice for every $\alpha$, including the 3Q state (Fig.~\ref{fig:theo_MAE_ASE}(b)). For a given sign of the ASE, the energy difference between the different superposition states, i.e.\ a variation of $\alpha$, is small with only up to 0.2 meV/Mn atom. In contrast, the energy difference between different spin configurations for the same $\alpha$ are much larger, on the order of up to 1 meV/Mn atom. We have also studied the effect of magnetic dipole-dipole interactions on the different states, however, the energy scale is much lower with a maximum energy difference of 0.25 meV/Mn atom~\cite{supplmat}. We find that for a 3Q ground state the ASE always favors the 3Q$^{1}$ orientation, regardless of the sign of the ASE. The sign of the ASE selects the specific orientation of the spins within the 3Q$^{1}$ state. 

The two configurations corresponding to lowest energy states for the different signs of the ASE are displayed in Fig.~\ref{fig:theo_MAE_ASE}(c) and differ only by a rotation around the surface normal. For a negative sign of the ASE the 3Q$^{1}_{\textrm{M}}$ state is favored and for a positive sign the 3Q$^{1}_{\textrm{K}}$ state. For each of these 3Q$^{1}$ states the energy is invariant under the inversion of all spins (e.g.~from all-out to all-in) and under the inversion of the $z$-components. This leads to four degenerate states \cite{supplmat}, which is confirmed by self-consistent DFT calculations including SOC for the four 3Q$^{1}_{\textrm{M}}$ states and one 3Q$^{1}_{\textrm{K}}$ state for Pd/Mn/Re(0001). The obtained total energy difference between the 3Q$^{1}_{\textrm{M}}$ states is negligible and the energy difference $\Delta E$ between the 3Q$^{1}_{\textrm{M}}$ and 3Q$^{1}_{\textrm{K}}$ state amounts to $0.75 \, \text{meV}$ per Mn atom~\cite{supplmat}, in very good agreement with the results obtained by the atomistic model (Fig.~\ref{fig:theo_MAE_ASE}(b,c)) with the ASE constant determined from DFT.

In conclusion, we have inspected the different energy terms that govern the detailed spin configuration of the 3Q state. We find that in general the ASE determines both the orientation and the relative spin alignments of the ideal 3Q state. In the case of Pd/Mn/Re(0001), it is the 3Q$^{1}_{\textrm{M}}$ configuration which exhibits three-fold symmetry and therefore does not show rotational domains, in agreement with the experimental data. Distorted 3Q states are always of 3Q$^2$ or 3Q$^3$ type, depending on the sign of MAE, and can be identified in experiments by the presence of rotational domains according to their two-fold symmetry on a triangular lattice. In general, for distorted states both the topological orbital moment and the topological Hall effect will be reduced compared to the ideal 3Q state. Our work provides key microscopic insight into the detailed spin configuration of the 3Q state. The knowledge of the magnetic moment directions is important for an understanding of the response of domains and domain walls in such topological orbital ferromagnets to external magnetic fields, which in turn is crucial for their transport properties.

\begin{acknowledgments}
We thank Moritz Goerzen for valuable discussions. K.v.B.\ and S.H.\ gratefully acknowledge financial support from the Deutsche Forschungsgemeinschaft (DFG, German Research Foundation) via projects no.~402843438, no.~418425860, and no.~462602351, and computing time provided by the North-German Supercomputing Alliance (HLRN).
\end{acknowledgments}

\end{document}